\documentclass[12pt]{iopart}
\usepackage{epsfig}
\begin{document}

\begin{flushright}
INT--PUB 04--07
\end{flushright}
\vspace{0.1in}

\title{The Color Glass Condensate at RHIC}

\author{Jamal Jalilian-Marian}

\address{Institute for Nuclear Theory, University of Washington, 
Seattle, WA 98195 USA}

\begin{abstract}
The Color Glass Condensate formalism and its application to high energy heavy 
ion collisions at RHIC are discussed. We argue that the RHIC data supports 
the view that the Color Glass Condensate provides the initial conditions for 
gold-gold collisions at RHIC while final state (Quark Gluon Plasma)
effects are responsible for the high $p_t$ suppression in mid rapidity. At forward 
rapidities in deuteron-gold collisions, however, Color Glass Condensate is the 
underlying physics of the observed suppression of the particle spectra and 
their centrality dependence.

\end{abstract}




 \ead{jamal@phys.washington.edu}

\section{Introduction}

The recent data from forward rapidity deuteron-gold collisions at the Relativistic Heavy 
Ion Collider (RHIC) in Brookhaven National Laboratory (BNL) has generated much interest 
and has led to intense debate on the underlying physics of the forward rapidity processes 
in deuteron-gold collisions. While the conventional models of multiple scattering based 
on Glauber theory predicted an enhancement of the Cronin effect in the forward rapidity 
region, the data shows a clear suppression of the negatively charged hadron spectra, as 
compared to the normalized proton-proton collisions, in agreement with the predictions 
based on the Color Glass Condensate formalism.  The most recent data on centrality 
dependence of the suppression in the forward rapidity region also agrees with the predictions 
of the Color Glass Condensate formalism and is in clear contradiction with expectations 
based on the conventional models.

The Color Glass Condensate formalism is an effective theory of QCD at high energy 
(small $x$). At high energy or equivalently (for fixed $Q^2$) at small $x$, gluons 
are the most abundant partons. While one can not calculate parton distribution 
functions in QCD using perturbation theory, one 
can calculate their evolution with $x$ or $Q^2$ in pQCD. The evolution of the 
gluon distribution function is described by the DGLAP evolution equation which sums 
$\alpha_s \log Q^2$ type corrections to the gluon distribution function. In DGLAP 
formalism, the gluon distribution function $xG(x,Q^2)$ grows fast with both $x$ and 
$Q^2$ as shown in Eq. (\ref{eq:dglap}) for fixed a coupling constant $\alpha_s$. 
It is worth noting that DGLAP evolution equation is just the renormalization 
group equation for the parton number operator.

\begin{eqnarray}
xG(x,Q^2) \sim \exp\bigg[\sqrt{\alpha_s \log 1/x \log Q^2}\bigg]
\label{eq:dglap}
\end{eqnarray}

If the phase space in $Q^2$ is limited and there is a large phase space available in $x$, then
BFKL is the applicable formalism. In BFKL approach, one sums terms of the form $\alpha_s \log 1/x$
which leads to an even faster grows of the gluon distribution function with $x$ (energy) compared 
to the DGLAP gluons

\begin{eqnarray}
xG(x,Q^2) \sim e^{\lambda \log 1/x}
\label{bfkl}
\end{eqnarray}
with $\lambda \equiv {N_c \alpha_s \over \pi}4\, \log 2$ for a fixed coupling constant. 
In other words, pQCD radiation leads to a fast growth of the gluon distribution function 
with energy or $x$ which makes a hadron a dense system of gluons at high energy. The 
gluon density in a high energy nucleus is even larger by a factor of $\sim A^{1/3}$ 
due to the Lorenz contraction of the longitudinal size of the nucleus at high energy. 
However, this fast growth can not go on forever since it would lead to violation of 
unitarity for physical cross sections.

Both DGLAP and BFKL approaches are linear in the gluon density. In other words, they only 
include radiation of gluons. As the phase space density of gluons increases, it becomes as
 probable for a gluon at small $x$ to recombine with another gluon at small $x$ and become 
a gluon at a higher $x$ as it is for a gluon at larger $x$ to radiate another gluon at a 
smaller $x$. This recombination of gluons slows down the growth of the gluon distribution 
function and leads to its eventual unitarization. This was first investigated by Gribov, 
Levin and Ryskin (GLR) and later on by Mueller and Qiu (MQ). The evolution equation for 
the gluon distribution function is written as (in the Double Logarithm Approximation)
\begin{equation}
{\partial^2 xG(x,Q^2) \over \partial \log 1/x \, \partial \log Q^2}  = 
{N_c \alpha_s \over \pi} xG(x,Q^2)
- {4\pi^3 \over (N_c^2 -1)} ({N_c\alpha_s \over \pi})^2 {1 \over Q^2} x^2 
\, G^2 (x, Q^2)
\label{eq:glrmq}
\end{equation}
Here $G^2(x,Q^2)$ is the four point function of gluons while the standard gluon distribution 
function $xG(x,Q^2)$ is the two point function of gluons. It is quite customary to factorize 
the four point function $G^2$ in terms of the gluon distribution function $G$ as 
$G^2 (x,Q^2)\sim {1 \over \pi R^2} [G(x,Q^2)]^2$ where $R$ is a scale which can not be 
determined from pQCD. This factorization is an approximation where one ignore gluon correlations, 
valid at large $N_c$. The fact that this is a higher twist correction to the gluon distribution 
function is manifested in the presence of the $1/Q^2$ factor in the non linear correction.

As the phase space density of gluons becomes large, even the GLR-MQ approach breaks down since 
all higher point functions become as large as the two point function and one must keep all 
higher order terms in the evolution equation. The Color Glass Condensate is an effective 
field theory approach to QCD at high energy (small $x$) which extends the applicability of 
pQCD to a dense system of gluons. 

\section{The Color Glass Condensate}

\begin{figure}[htp]
\centering
\setlength{\epsfxsize=12cm}
\centerline{\epsffile{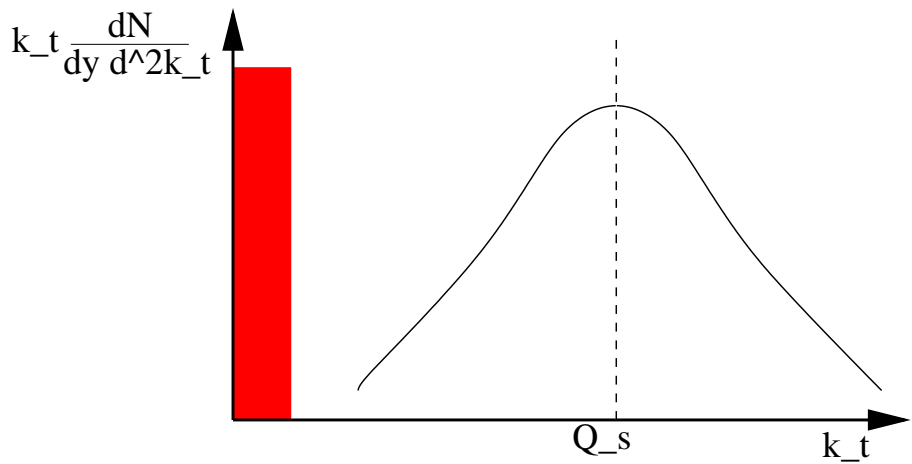}}
\caption{Phase space distribution of gluons}
\label{fig:kdndk}
\end{figure}

The effective action for QCD at small $x$ is given by (for a recent review, see \cite{edraj})
\begin{eqnarray}
S&=&i\int d^2 x_t F[\rho ^a(x_t)]\nonumber \\
&-& \int d^4 x {1\over 4}G^2 + {{i}\over{N_c}} \int d^2 x_t dx^-
\delta (x^-)
\rho^{a}(x_t) {\rm tr}T_a W_{-\infty,\infty} [A^-](x^-,x_t)
\label{eq:action}
\end{eqnarray}
where $G^{\mu \nu} $ is the gluon field strength tensor
\begin{eqnarray}
G^{\mu \nu}_{a} = \partial^{\mu} A^{\nu}_{a} 
- \partial^{\nu} A^{\mu}_{a} + 
g f_{abc} A^{\mu}_{b} A^{\nu}_{c}
\end{eqnarray}
$T_a$ are the $SU(N)$ color matrics in the adjoint representation,
and $W$ is the path ordered exponential along the $x^+$ direction in
the adjoint representation of the $SU(N_c)$ group
\begin{equation}
W_{-\infty,\infty}[A^-](x^-,x_t) = P\exp \bigg[-ig \int dx^+
A^-_a(x^-,x_t)T_a \bigg]
\end{equation}
In order to calculate a physical quantity, one averages over the gluonic field configuration 
\begin{equation} 
<O>={\int [D\rho^a][DA^\mu_a]O(A)\
\exp\{iS[\rho, A]\}\over \int [D\rho^a][DA^\mu_a] 
\exp\{iS[\rho, A]\}}
\nonumber
\label{eq:average}
\end{equation}
The weight function 
\begin{equation}
Z\equiv \exp\{-\int d^2x_t F[\rho^a]\}
\nonumber
\label{eq:bolz}
\end{equation}
appearing in (\ref{eq:average}) is the statistical weight of a particular configuration of the 
two dimensional color charge density $\rho^a(x_t)$ inside the hadron. Initially, this statistical 
weight is taken to be a Gaussian
\begin{equation}
F[\rho(x_t)]= {{1}\over {2\mu^2}} \rho_a^{2}(x_t)
\nonumber
\label{eq:gaussian}
\end{equation}
The Gaussian approximation is valid as long as the color charge density is large and random.
Also, if one is interested in the gluon correlation at the same rapidity, one needs to include the 
longitudinal structure of the sources into account. This is done in \cite{jkmw} where one solves
the classical equations of motion in the presence of sources which have a longitudinal size. The
result for the two point function is 
\begin{eqnarray}
    G_{ij}(y,x_\perp;y^\prime,x_\perp^\prime) = 
\langle A_i(y,x_\perp) A_j(y^\prime,x_\perp^\prime)\rangle
\end{eqnarray}
Performing the color averaging with the extended sources, we get
\begin{equation}
G_{ii}^{aa}=\frac{4(N_c^2-1)}{N_c x_\perp^2}
     \left[1-\left(x_\perp^2\Lambda_\mathrm{QCD}^2\right)^{
     \frac{g^4N_c}
      {8\pi}\chi(y)x_\perp^2}\right]
\label{eq:propagator}
\end{equation}
The momentum space distribution of the gluon distribution function obtained from the Fourier 
transform of Eq. (\ref{eq:propagator}) has a much milder divergence at low $k_t$ than the 
pQCD distributions due to the non linearities of the gluons included in a classical description. 
One can also show that gluons typically have a transverse momentum $Q_s (x,b_t,A)$, called the 
saturation scale, which can be much larger than $\Lambda_{QCD}$. This is shown in Fig. 
(\ref{fig:kdndk}) which illustrates the fact that most gluons reside in a state with a momentum 
$\sim Q_s$ described by a strong classical field.
As one goes to smaller $x$, one needs to include quantum evolution effects, namely, sum the 
$\alpha_s \log 1/x$ terms which become large. This has been done and leads to 
a non linear functional equation for the weight functional $F(\rho)$ which can then be used 
to write an evolution equation for any desired physical observable. The JIMWLK equation for 
the weight functional \cite{jimwlk} can be written in a compact form as (first written in this 
form by H. Weigert)
\begin{eqnarray}
\partial_y Z_y[U] = - {1 \over 2}\, i \, \nabla^a_{x_t} \, \chi^{ab}_{x_t\,y_t}\,i\,\nabla^b_{y_t} 
Z_y[U]
\label{jimwlk}
\end{eqnarray}
with $y\equiv \log 1/x$. These equations are highly non linear and coupled so that one has to 
develop approximate techniques to solve them. They have also been investigated numerically. 
One particular limit which simplifies the non linear JIMWLK equations, is the large 
$N_c$ limit which was used in \cite{yk} to derive an equation for the dipole scattering cross 
section. This equation is known as the BK equation and is widely regarded as the simplest 
possible non linear equation which respects perturbative unitarity. The BK equation is 
simplest in the momentum space where it can be written as 
\begin{eqnarray}
{\partial \bar{N} (x,k_t,b_t) \over \partial y} = 2 \bar{\alpha} \chi (- {\partial \over 
\partial \log k}) 
\bar{N}(x,k_t,b_t) - \bar{\alpha} \bar{N}^2(x,k_t,b_t)
\end{eqnarray}
where $\bar{N}$ is related to the dipole cross section via 
\begin{eqnarray}
\bar{N}(x,k_t,b_t) \equiv \int {d^2 r_t \over 2\pi r_t^2} e^{-ik_t\cdot r_t} \sigma (x,r_t,b_t)
\end{eqnarray}
and the dipole cross section $\sigma (x,r_t,b_t)$ is defined in terms of the expectation 
value of the correlator of two Wilson lines
\begin{eqnarray}
\sigma (x,r_t,b_t) \equiv {1 \over N_c} Tr < 1 - U(x_t) U^{\dagger}(y_t) >
\label{eq:sig_dipole}
\end{eqnarray}
with $r_t = x_t - y_t$ and $b_t= (x_t + y_t)/2$. The BK equation has been investigated both 
numerically and analytically and approximate analytic solutions as well as complete numerical 
solutions have become available.

\section{Applications to RHIC}

\begin{figure}[htp]
\centering
\setlength{\epsfxsize=8cm}
\centerline{\epsffile{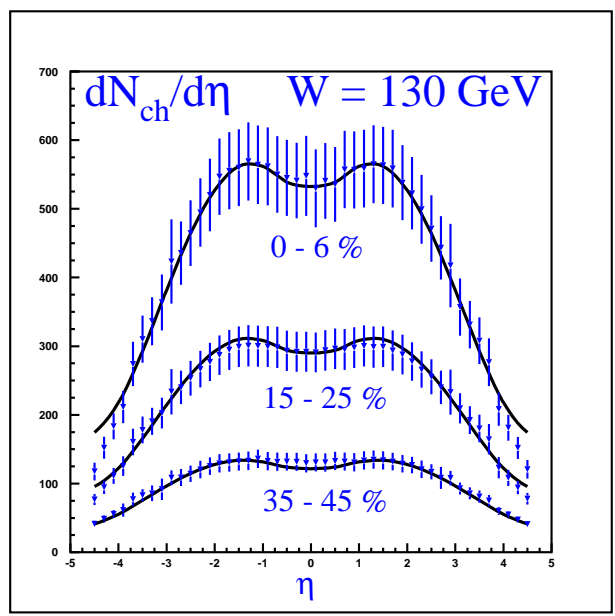}}
\caption{Hadron multiplicity}
\label{fig:kln_mul}
\end{figure}

The Color Glass Condensate was first applied at RHIC to the measured hadron multiplicities in 
nucleus-nucleus collisions at $\sqrt{s}=130\, GeV$ \cite{kln}. The starting point is the $k_t$ 
factorized form of gluon production cross section in heavy ion collisions
\begin{eqnarray}
E{d\sigma\over d^3p} = {4\pi N_c\over N_c^2 - 1} {1\over p_t^2} \int dk_t^2 \,\alpha_s\,
\phi (x_1,k_t^2) \phi (x_2, (p_t-k_t)^2)
\label{eq:kt_fact}
\end{eqnarray}
where $x_1,x_2={p_t\over\sqrt{s}}\exp(\pm \eta)$. To get particle multiplicities, one 
integrates (\ref{eq:kt_fact}) over the transverse momentum and divides by the inelastic 
cross section of nucleus-nucleus collisions. The result is shown in Fig. (\ref{fig:kln_mul}) 
where a good agreement with the data at $\sqrt{s}=130\,GeV$ is obtained. The centrality and 
energy dependence are also in good agreement with the RHIC data. 
It should be noted that 
{\it multiplicities are dominated by low transverse momenta}. For example, more than $95\%$ 
of produced particles are produced with transverse momentum less than $1\, GeV$ which 
indicates that multiplicities probe  $x_1,x_2$ of the nucleus wave functions which are 
smaller than  $0.01$ at mid rapidity.

Eq. (\ref{eq:kt_fact}) was also applied to hadron transverse momentum spectrum in mid rapidity 
nucleus-nucleus collisions at $\sqrt{s}=130\, GeV$ and a reasonable agreement was obtained. This 
would indicate that the observed suppression of the hadron spectra in mid rapidity nucleus-nucleus 
collisions is due to initial state (and not Quark Gluon Plasma) effects. Deuteron-gold collisions 
were performed in order to distinguish initial state from final state effects. The observed back 
to back hadron correlations and a lack of suppression in hadron spectra in mid rapidity 
deuteron-gold collisions
basically rule out high gluon density effects at high $p_t$ in mid rapidity gold-gold collisions and 
are a strong indication of formation of a dense, strongly interacting medium which is most likely 
partonic in origin. As such, the Color Glass Condensate provides the initial conditions for the 
formation of the Quark Gluon Plasma \cite{hn}. However, the physics of mid rapidity nucleus-nucleus 
collisions is mostly final state interactions between produced partons. Therefore, it is not the 
best place to look for manifestation of the Color Glass Condensate because, at high $p_t$, the 
final state effects in mid rapidity nucleus-nucleus collisions are clearly more important than 
initial state effects. 

On the other hand, the forward rapidity region in deuteron-gold collisions is mostly free of 
these final state interactions since there is no Quark Gluon Plasma expected to be formed there.
Also, in the forward rapidity region in deuteron-gold collisions, one probes the smallest $x$ 
in the nucleus wave function kinematically possible. This is where the Color Glass Condensate 
effects are expected to be the strongest. Another reason why the forward rapidity region is the 
best place to look for the Color Glass Condensate effects in a collider environment is that in 
forward rapidity, the experimental $p_t$ coverage is limited by the kinematics. Therefore, 
evolution in $p_t$ is not very important while the effects of $x$ evolution are maximized. 
This is crucial since the Color Glass Condensate formalism does not include $\alpha_s\log Q^2$ 
effects (summed by DGLAP) which become important at high $p_t$ (the double log region where one 
sums $\alpha_s \log Q^2 \log 1/x$ terms is indeed included in CGC).

\begin{figure}[htp]
\centering
\setlength{\epsfxsize=8cm}
\centerline{\epsffile{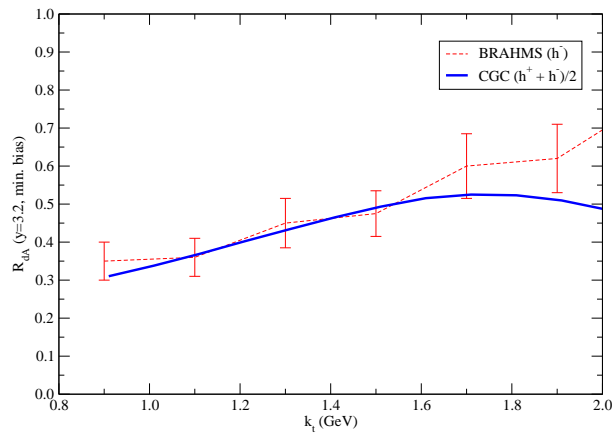}}
\caption{Minimum bias $R_{dA}$ for forward rapidity ($y=3.2$) deuteron-gold collisions.}
\label{fig:R_dA_32}
\end{figure}

In the forward rapidity region of deuteron-gold collisions one probes the large $x$ part of the 
deuteron wave function. To simplify the discussion, we will focus on proton-nucleus collisions. 
In this kinematic region, proton is a dilute system of partons with their distributions well 
described by pQCD. On the other hand, the nucleus is probed at very small $x$ so therefore, 
we treat it as a Color Glass Condensate. 
The starting point is to consider scattering of a valence quark of the proton on the nucleus 
\cite{djm}. One could also consider radiation of a gluon from the quark scattering on the nucleus 
\cite{km}. The properties of these cross sections have been studied in detail in \cite{cron,kkt}
where it shown that Color Glass Condensate formalism, at the classical level, leads to the Cronin 
effect while inclusion of $x_{bj}$ evolution leads to a suppression of the nuclear enhancement factor.
Since the case when the valence quark scatters on the nucleus is simpler (and 
possibly the most important in the forward rapidity and low $p_t$), we consider scattering 
of a quark on the nucleus described as a Color Glass Condensate. The differential cross 
section is given by
\begin{eqnarray}
{d\sigma^{qA\to qX} \over  d^2 b_t\, d^2 q_t\,dq^-} =\frac{1}{(2\pi)^2} \delta (p^- - q^-)
\int d^2 r_t e^{i q_t \cdot r_t} \sigma (x,r_t,b_t)
\label{eq:qA}
\end{eqnarray}
where the dipole cross section $\sigma (x,r_t,b_t)$ is given by (\ref{eq:sig_dipole}). It is 
interesting to note that this cross section is finite as $q_t \rightarrow 0$ unlike the pQCD 
cross sections which are divergent. 
To obtain cross sections for hadron production, one needs to convolute (\ref{eq:qA}) with the 
quark distribution function in a proton and a quark-hadron fragmentation function
\begin{eqnarray}
{d\sigma^{pA\to \pi(k)X} \over dk^-d^2k_t}\equiv
\int dx_q\, dz\, q_p(x_q)\,\,
{d\sigma^{qA\to qX} \over dq^-d^2 q_t} \,\,
D_{q/\pi}(z)
\label{eq:fac}
\end{eqnarray}
In Fig. (\ref{fig:R_dA_32}) we show the result of a recent calculation \cite{jjm} based on 
(\ref{eq:fac}) for minimum biased deuteron-gold collisions in forward rapidity ($y=3.2$) 
and compare it with the nuclear modification factor $R_{dA}$ for negatively charged hadrons 
measured by BRAHMS \cite{brahms} (our calculation is for the sum of positively and negatively 
charged hadrons, 
but in the kinematic region considered here, this does not make a big difference). The agreement 
at low $p_t$ is quite striking specially since there is absolutely no free parameter used here.

The physics of particle production in forward rapidity and low $p_t$  can be understood in 
simple terms.
The valence quark of the deuteron scatters from the fully developed (as allowed by the kinematics) 
wave function of the nucleus, characterized by $Q_s^A (y)=\sqrt{2} e^{\lambda y/2} \sim 2.2\, GeV$ 
(at $y=3.2$) which is roughly how much transverse momentum the scattered quark accumulates. Assuming 
$<z> \sim 0.8$ in the fragmentation process, this means quark-nucleus scattering can describe 
particle production up to $\sim 1.8 \, GeV$. As one goes to higher $p_t$, one needs to include
 radiation of gluons in analogy with DGLAP radiation. In \cite{km}, gluon radiation from a 
quark scattering on the nucleus is calculated and must be included at higher $p_t$. Indeed, 
the suppression of hadron multiplicities in deuteron-gold collisions in the forward rapidity 
region had been predicted \cite{kkt} based on the behavior of the gluon production 
cross section in different $p_t$ regions as was the striking centrality dependence of the 
hadron spectra in different rapidity regions. The fact that the cross section for scattering 
of quarks on the nucleus has similar properties is not surprising since both quark scattering 
and gluon production cross section depend on the (fundamental vs. adjoint) dipole cross section.

\section{Summary}

The recent results from the Relativistic Heavy Ion Collider have provided much excitement and 
enthusiasm in the high energy heavy ion community. On the one hand, there is overwhelming
evidence for the formation of a deconfined state of matter, most likely of partonic nature 
in thermal equilibrium, in mid rapidity heavy ion collisions. On the other hand, we have, 
very likely, seen the first unambiguous evidence for the Color Glass Condensate in the forward 
rapidity deuteron-gold collisions. Even though the Color Glass Condensate correctly predicted 
rapidity, energy and centrality dependence of hadron multiplicities in gold-gold collisions, 
the more conventional models, based on Glauber multiple scattering, were also able to accommodate
the data. 

The spectacular failure of the conventional models applied to the forward rapidity region in
deuteron-gold collisions and the agreement of the Color Glass Condensate predictions with the 
observed suppression of the nuclear modification factor and its centrality dependence have 
provided strong evidence in favor of the Color Glass Condensate formalism. To put this on a more
firm footing experimentally, one needs to measure more observables, such as photons, 
dileptons, etc. \cite{fg,bms}. 
Another deuteron-gold run at RHIC will go a long way towards establishing the properties of the 
Color Glass Condensate and is urgently needed.

\section{Acknowledgement}

This work is supported by DOE under grant number DOE/ER/41132.

\end{document}